# Tuning Optical Conductivity of Large-Scale CVD Graphene by Strain Engineering

*Guang-Xin Ni[1,2,3], Hong-Zhi Yang[1], Wei Ji[1,2], Seung-Jae Baeck[4], Chee-Tat Toh[1,2], Jong-Hyun Ahn[4], Vitor M. Pereira[1,2]\*, and Barbaros Özyilmaz[1,2,3,5]\**

[1]Department of Physics, 2 Science Drive 3, National University of Singapore, Singapore 117542
[2]Graphene Research Centre, 6 Science Drive 2, National University of Singapore, Singapore 117546
[3]NanoCore, 4 Engineering Drive 3, National University of Singapore, Singapore 117576
[4]School of Electrical & Electronic Engineering, Yonsei University, Seoul 120-749, Korea
[5]NUS Graduate School for Integrative Sciences and Engineering (NGS), Singapore 117456
*Corresponding Authors: barbaros@nus.edu.sg; vpereira@nus.edu.sg

ABSTRACT Strain engineering has been recently recognized as an effective way to tailor the electrical properties of graphene. In the optical domain, effects such as strain-induced anisotropic absorption add an appealing functionality to graphene, opening the prospect for atomically thin optical elements. Indeed, graphene is currently one of the notable players in the intense drive towards bendable, thin, and portable electronic displays, where its intrinsically metallic, optically transparent, and mechanically robust nature are major advantages. Given that the intrinsic transparency of a graphene monolayer is 97.7 %, any small, reproducible, controllable, and potentially reversible modulation of transparency can have a significant impact for graphene as a viable transparent conducting electrode. Even more so, if the degree of modulation is polarization dependent. Here we show that the transparency in the visible range of graphene pre-strained on a Polyethylene terephthalate (PET) substrate exhibits a periodic modulation (0.1 %) as a function of polarization direction, which we interpret as strain-induced optical anisotropy. The degree of anisotropy is varied by reversible external manipulation of the level of pre-strain. The magnitude of strain is monitored independently by optical absorption and Raman spectroscopy, and the experimental observations are consistent with the theoretically expected modification of the optical conductivity of graphene arising from the strain-induced changes in the electronic dispersion of graphene. The strain sensitivity of the optical response of graphene demonstrated in this study can be potentially utilized towards novel ultra-thin optical devices and strain sensing applications.

Graphene keeps attracting much attention, and continues the subject of enormous amounts of experimental and theoretical activity since its first micromechanical exfoliation in 2004 [1-4]. Being an atomically thin membrane, graphene's electrical, optical, and chemical properties are highly amenable to external influences, including externally induced doping, surface and lattice defect engineering, selective substrate interactions, or strain engineering [5-9]. Strain engineering, in particular, is appealing since it potentially allows incremental and reversible control over various of its technologically-critical properties, such as electronic transport [10], optical absorption [11], chemical affinity [12], or bio-compatibility [13]. Conversely, the response of these properties to lattice deformations can be explored in strain sensing scenarios where an ultrathin, transparent or flexible sensor would be desirable [14]. This prospect is appealing in graphene because, despite its atomic thickness, it can withstand reversible mechanical deformations in excess of 15% [15, 16]. Adding to this unparalleled mechanical robustness, the honeycomb structure of the underlying crystal lattice leads to an unusual coupling between deformations and electrons. Electrons tend to feel planar deformations as an effective magnetic field, in addition to the conventional displacement field coupling that arises from the change in unit cell volume [17]. While the latter acts essentially as a local electrostatic potential modulation, and can therefore be effectively screened by the free electrons, the pseudo-magnetic field nature of the coupling in graphene has a stronger impact on the electronic structure. This has been partially confirmed recently in a striking fashion, with experimental reports of Landau levels in graphene corresponding to strain-induced pseudo-magnetic fields in the unheard of range between 300 and 600 T [18, 19]. This documents the peculiar nature of the interplay between deformations and electronic structure in graphene, and shows that strain can indeed strongly impact the electronic properties of this system in a continuous and tunable way.

Equally remarkable is the optical response of graphene. The most striking feature is its constant and universal optical absorption over a broad band of frequencies which, for undoped systems, spans the IR to visible range [20], and is tunable via electronic doping by virtue of the Pauli blocking mechanism below a density-dependent cut-off frequency [21]. Given that the optical response of graphene up to the UV is determined solely by its conduction electrons, strain-induced modifications of the electronic structure are bound to impact the optical response as well.

Under mechanical deformation, the variation of the carbon-carbon interatomic distance along the various directions modifies the electronic band structure of graphene, and could ultimately lead to a Lifshitz transition, upon which the topology of the band structure near the Fermi level changes from the characteristic Dirac nature to a gapped spectrum [22]. But such transition would require an amount of strain too close to the breaking point of graphene to be realistically possible [23]. At realistic and predominantly uniaxial deformations, the dominant modifications of the electronic structure can be captured by two factors: (i) the displacement of the Dirac point from the corners of the Brillouin zone [24], and (ii) the evolution of the low energy 2D band dispersion from an isotropic to an elliptical cone [11, 25]. Whereas the displacement of the Dirac point lies at the origin of the effective pseudo-magnetic field felt by the electrons in strained graphene, the anisotropy in the Fermi surface introduces two orthogonal, fast and slow, directions, which lead to two principal directions for minimal and maximal light absorption. The transmittance of linearly polarized light impinging perpendicularly to strained graphene is therefore expected to be modulated as the polarization direction varies with respect to the principal directions [11]. Various other related consequences have been theoretically explored [26, 27] but, to the best of our knowledge, a direct experimental observation and quantification of strain-induced changes in the optical absorption spectrum is still absent.

In this letter, we report the observation of a clear, controllable, reproducible and reversible modulation of optical transmissivity in large-scale graphene grown by CVD. The modulation is a result of optical anisotropy induced in the graphene sheet by pre-strain that is established during the CVD growth process and survives the transfer to a flexible and transparent PET film. Flexion of the graphene/PET structure results in weaker optical anisotropy, which is interpreted as the release of pre-existing compression in the transferred graphene. The clear periodic dependence of the transmissivity on the polarization of incoming light indicates that the transferred sheet is under predominantly uniaxial strain. This behaviour follows the theoretical expectation for the modulation of the optical conductivity under uniaxial strain. The optical absorption measurements are complemented by an analysis of the Raman shift which is consistent with this picture. Strain magnitudes extracted independently from the Raman and optical absorption measurements tally with each other.

In this study large-scale graphene was synthesized by CVD on pure copper foils [28]. By controlling the post-growth annealing time, one can synthesize predominantly monolayer graphene with coverage of up to 95%. Recent work [29] has shown that graphene as grown on Cu is under strain, which results from a combination of lattice mismatch between the two crystals, opposite signs of the thermal expansion coefficient, and the natural tendency for graphene to physisorb to the metal surface. Growth and annealing conditions are thus conducive to built-in strains in graphene at the end of the growth process. In the case of graphene grown on Cu single crystals, the nature of the strain distribution in the as-grown graphene is apparently (and not unexpectedly) highly dependent on the particular face where deposition occurs. For example, reference [29] reports that, when grown on a (111) face, graphene is biaxially compressed by ~0.5% and displays an isotropic Moire pattern under STM, whereas growth on a (100) face leads to more unidirectional features. In the case of Cu foils the strain distribution is more complex, and highly variable due to the polycrystalline nature of the metal. What is remarkable is that the state of strain can largely survive after the wetting transfer to a PET substrate, which is similar to the nano-printed graphene surface features, i.e., nanoripples, wrinkles after transfer and can be used to our advantage [6, 29]. Hence, if on the one hand this pre-existing strain in CVD graphene can be a limiting factor for the growth of large high-quality atomic layers, on the other it provides a convenient and versatile platform for the study of strain-induced modifications of some of graphene's intrinsic properties.

Subsequently to deposition, graphene films were transferred onto polyethylene terephthalate (PET) substrates using the transfer method introduced by Li *et al.* [30] (**Figure. 1**a). We resorted to a PET membrane as substrate due to its excellent flexibility and thickness control, which allows us to controllably and reproducibly vary the strain conditions of graphene. Moreover, PET is highly transparent, which permits efficient optical probing of the transmittance of graphene. Systematic atomic force microscope (AFM) studies on centimetre size samples reveal that the surface morphology of the PET is characterized by a high density of self-assembled protrusions, or nanopillars, with no particular spatial arrangement, or length scale (**Figure. 1**b). These protrusions might play a role in stabilizing the state of strain during and after the graphene transfer, but the specific nature of this process requires further scrutiny. Indeed, if control over the distribution of such PET protrusions can be achieved, this would provide further versatility in the level of strain manipulation, in the same vein as reported recently in graphene over a nanostructured $SiO_2$ substrate [31]. Raman scans over the same areas imaged by AFM are shown in Figs. 1c and 1d, and it is visible that the strain distribution is not homogeneous since the frequency of the Raman 2D peak fluctuates noticeably.

**Figure 2** shows the optical setup for the optical transmission measurements, from where we extract the polarization dependence of the transparency (see SI for a more detailed description). To minimize laser intensity fluctuations and shot noise level, a coherent detecting technique containing two identical ultra-high sensitivity optical detectors was utilized in our optical setup (**Figure. 2**a). **Figure 2**b shows the optical absorption of a graphene sheet on flat PET as a function of light polarization, and at optical frequencies (632.8 nm). Instead of the universal and polarization-independent transmission characteristic of isolated graphene, we observe a clear periodic modulation as a function of polarization angle. The different curves in **Figure. 2**b correspond to transmission measured at different regions of the sample, thus showing that the effect is independent of the location of the laser spot within the sample. The fact that the transmittance displays this clean periodic modulation strongly hints at a state of uniaxial strain [11]. To be more quantitative, the infrared optical conductivity of anisotropically strained graphene is expected to vary with the strain magnitude, $\varepsilon$, according to $\sigma_{T,L}(\omega) \approx \sigma_{iso}(\omega) [1\pm\lambda a(1+\nu)\varepsilon]$, where $\sigma_L$ ($\sigma_T$) represents the optical conductivity along the direction parallel/longitudinal (perpendicular/tranverse) to the extension direction, $\sigma_{iso}(\omega)$ is the conductivity in the isotropic case, $\nu$ the Poisson ratio, and $\lambda a$ is a numerical constant $\approx$ 3-4 [11]. As a result, the transmittance is predicted to display a modulation with respect to the polarization direction of the incoming light as $T \approx 1-\pi\alpha[1-\lambda a(1+\nu)\varepsilon \cos2\phi_I]$, with $\alpha$ being the fine structure constant, and $\phi_I$ the angle between the polarized electric field and the extension direction. The modulation amplitude is then expected to be $\Delta T \approx 2\pi\alpha\lambda a(1+\nu)\varepsilon$. Our data in **Figure. 2**b shows a modulation amplitude in the transparency of $\Delta T \approx 0.1\%$. If the Poisson ratio of ideal graphene ($\nu\approx0.16$) is used, this would correspond to a uniaxial strain of the order of 0.5%, on the basis of the estimate above. However, since the optical absorption is seen here to be anisotropic prior to any external tension, it implies that graphene is pre-strained, and thus the effective or relevant Poisson ratio can be rather different from the one in isolated graphene. Nevertheless it is notable that the just transferred graphene sheet displays a degree of optical anisotropy suggesting a pre-strain of the same magnitude known to exist in the graphene/Cu system at the end of the growth process [29]. In order to unequivocally ascribe the transmittance modulation to graphene alone, one has to rule out the contributions from the substrate. This has been addressed by measuring the transparency in regions free of graphene, which is seen to be constant, independent of the polarization direction (see SI for a more detailed account), and therefore confirms that the modulation with polarization direction is due solely to the anisotropic optical absorption of graphene.

To modify the state of strain in graphene, the graphene/PET structure was mounted on the mechanical setup shown in **Figure. 3**a, designed for controlled bending. The curvature was increased in various steps, with the polarization-dependent optical absorption and Raman spectrum measured at the apex of the bent structure with each increment. The finite thickness (170 μm) of the substrate means that curvature will directly translate to extension of the top surface that holds graphene relative to the neutral position. One thus expects the top face of the substrate to be under uniaxial tension along the bending direction with respect to the flat configuration. In each of the images in **Figure. 3**a we include the nominal strain *on the PET substrate*. This nominal strain imposed on the substrate is used in what follows to label and identify measurements taken at the corresponding curvature. It should be taken as a reference only, as it is not related to the actual strain in graphene. The results of the optical absorption and Raman shift measurements are summarized in **Figure. 3**b and 4. The first important detail is that the anisotropic optical absorption in the flat configuration decreases gradually as the curvature of the substrate increases. This is seen by the decrease of the modulation amplitude in **Figure. 3**b, which is interpreted as the result of progressive release of built-in, and predominantly uniaxial, compression in the graphene sheet. At the maximally bent configuration anisotropy with respect to polarization direction is still seen in the transparency of graphene.

Raman spectroscopy provides an accurate probe of strain in graphene due to the strong sensitivity of the main Raman peaks to lattice deformations [32]. The Raman spectra in **Figure. 4**a confirm the progressive release of compression in graphene as the structure is bent. The spectra were taken with laser power below 1 mW, and a spot size on the sample of ~1 μm. Overall, more than 20 different locations covering an area of more than 10000 μm$^2$ have been scanned (see SI). Due to a strong and dominant Raman feature of the PET substrate overlapping with the G peak, we concentrate our analysis in the behaviour of the 2D peak only. **Figure. 4**b shows that the Raman 2D band appears at 2700 cm$^{-1}$ in the flat configuration, which corresponds to a blue-shift of 20 cm$^{-1}$ with respect to the frequency of pristine graphene at our excitation wavelength [33]. The presence of the blue-shifted Raman bands in the flat configuration is consistent with the interpretation of pre-existing compression in graphene transferred to PET, and the magnitude of the blue-shift is very close to the blue-shift measured in graphene as-grown on Cu foils [29], which, as in the case of the strain extracted from the optical anisotropy in **Figure. 2**b, hints at an effective preservation of the state of strain in graphene during transfer. We also highlight that the FWHM of the 2D band in these samples is considerably smaller than the one reported in [29] for CVD on Cu foils. This can be related to less strain inhomogeneity in our samples, and explain why the optical anisotropy is clearly seen here, despite the fact that graphene grown on Cu foil is expected to display strain non-uniformity. By gradually increasing the bending curvature, an obvious red-shift of the 2D peak was observed (**Figure. 4**a), down to ~2680 cm$^{-1}$ when the nominal external strain reaches ~0.7 %. This indicates that the pre-compressive strain is effectively released during the bending, which is consistent with the changes seen in the magnitude of the optical anisotropy. According to Ref. [32], in suspended graphene the rate of change in the 2D peak position with strain is expected to follow $\Delta\omega_{2D}/\Delta\varepsilon \sim$ -83 cm$^{-1}$/%. A down-shift of 20 cm$^{-1}$ would then correspond to a reduction in the compression of graphene of $\Delta\varepsilon$~0.25%. Even though we cannot ascertain how accurate this estimate is in our case due to the uncertainty in the magnitude of the Poisson effect in the graphene/PET structure, it is consistent with the magnitude of pre-existing compression extracted from the optical anisotropy: if the pre-existing compression is of the order of 0.5%, as extracted from the anisotropy in transmissivity, and if the maximally bent configuration still shows some remnant anisotropy, it is not unexpected that the maximally bent configuration releases most of the 0.5% compression, but not completely. Hence, the displacement of the Raman 2D peak should reveal a variation of strain close, but below, 0.5%, which is the case with the data shown in **Figure. 4**.

Finally, the Raman data also show that the initial pre-compressed state of graphene is recovered once the external strain (bending) is suppressed, and the effect is resilient to cycling. Further investigation is under way to assess the role of the PET nanopillar morphology, and its interaction with graphene, in allowing a noticeable transfer of strain from the Cu deposition stage to the final setup. Here we also point out that the small magnitude of pre-strain obtained in this study leads to correspondingly small modulations of the transmittance in absolute terms, and hence, despite the proof-of-principle and confirmation of the theoretical predictions for the magnitude of this modulation, is still far from those practical applications which require large optical modulations. Before conclusion, we would also like to mention another approach which is directly pattern graphene into different periodic features [34, 35]. Using this approach, potentially a voltage controlled anisotropic surface conductivity with large optical modulation would be achieved [34]. This approach stands for another efficient and easy-to-implement way with the capability to dynamically control the electromagenetic wave propagations, especpically in the THz and microwave frequencies [34].

In summary, we report a direct observation of controllable optical anisotropy in CVD graphene. Pre-strained graphene transferred to PET is seen here to exhibit a polarization dependent transparency, an effect that is interpreted as arising from the strain-induced changes in the electronic dispersion of graphene at low densities. The degree of anisotropy is proportional to the magnitude of non-isotropic strain and can therefore be controlled externally by mechanical means, for example. Further work along these lines is important and needed for the prospect of atomically thin optical elements for integration in optoelectronic and photonic devices.

**Experiemental Section**

Sample fabrication begins with large-scale graphene synthesized by CVD on pure copper foils, the details of graphene fabrication procedures are discussed in Refs. [28, 30]. The PMMA Cu-CVD graphene structure was immersed into copper etchant, and once the copper was etched away, the large-scale graphene was immediately transferred onto the PET substrates.

Optical measurements were performed under ambient conditions, in a dark space. A stable He-Ne (632.8 nm) laser was used with an intensity limited to less than 1 mW to avoid laser induced heating and was split by a beam splitter: the transmitted part was illuminated onto the graphene/PET structures which were mounted in the strain generator for the transmission measurements under bending; and the reflected part was used as a reference beam. The radius of the illuminated area on the graphene/PET was estimated to be ~ 200 μm. The polarization of the illuminating laser beam was controlled by rotating the λ/2 wave-plate. For the estimation of the nominal strain on the top surface of the PET substrate, we used $\varepsilon \approx d/2r$ for the calculations. Here d is the thickness of the PET film and r is the radius of the curvature.

Raman spectroscopy/imaging was carried out with a WITEC CRM200 Raman system with a 532 nm (2.33 eV) excitation frequency, and laser power at the sample below 1 mW to avoid laser-induced heating. A 50× objective lens with a NA=0.95 was used in the Raman experiments. Data analysis was done using WITec Project software.

**Acknowledgements.** The authors gratefully acknowledge the assistance from Yuda Ho in the device preparation. This work is supported by the Singapore National Research Foundation Fellowship Award NRF-RF2008-07, NRF-CRP6-2010-05, NRF-CRP Award "Novel 2D materials with tailored properties: beyond graphene" R-144-000-295-281, Singapore Millennium Foundation (SMF)-NUS Research Horizons Award 2009-Phase II, A*STAR SERC TSRP-iNPBi and the NUS Young Investigator Award.

Supporting Information is available online from Wiley InterScience or from the author.

Figure Captions

**Figure. 1**. (a) Optical image of our graphene/PET sample. The square shape with side length 2cm shows the CVD graphene film. (b) AFM image revealing the morphology of graphene on the PET substrate, dominated by various protrusions from the PET substrate. (c) Two-dimensional scan of the 2D Raman frequency. The scalebar is 2 μm, and the black areas correspond to the protruding substrate regions. (d) Variation of the 2D Raman frequency through a nanopillar, along the trace indicated by the dotted line in panel (c).

**Figure.2**. (a) Schematic of the setup for optical absorption measurements. (b) Optical transmittance of CVD graphene as a function of polarization for the flat substrate. A clear modulation of the optical absorption is observed, with a period of 180 degrees. The transmission changes from peak-to-peak by ≈ 0.1 %, which corresponds to a deformation of ~ 0.5% in graphene if a model of uniaxial strain is used. The different curves represent different locations of the laser spot within the graphene film.

**Figure. 3**. (a) View of the setup used to generate external strain. The geometry (radius of curvature and thickness) of the bent substrate is used to estimate the nominal strain on the top surface of the substrate. The corresponding nominal strain is indicated at the top of each panel. Note that the red dot is the indicator of the laser spot position. (b) Evolution of the optical transmittance with increasing curvature of the substrate. The progressive bending leads to a corresponding reduction in the modulation amplitude.

Figure. 4. (a) Evolution of the Raman spectrum of CVD graphene around the G and 2D bands during a bending-release cycle. The dash dot curves corresponding to the experimental data while the solid curves corresponding to the Lorentz fitting. The left portion is dominated by a PET feature around 1615 cm$^{-1}$ overlapping with the G band. The 2D peak is isolated from substrate features, and is seen to red-shift by ≈ 20 cm$^{-1}$ between the flat and maximally curved situation, which is evidence that the actual strain in graphene is reduced with bending. The symmetric recovery of the peak positions upon returning the substrate to the flat configuration indicates that the initial strain within the graphene lattice is restored. (b) The extracted $\omega_{2D}$ and $\omega_G$ vs. external strain. Here, the reason that G peak does not shift too much as a function strain is due to the pre-strain in our samples is not uniform, and a certain amount of biaxial strain might exist in our CVD graphene samples [29].

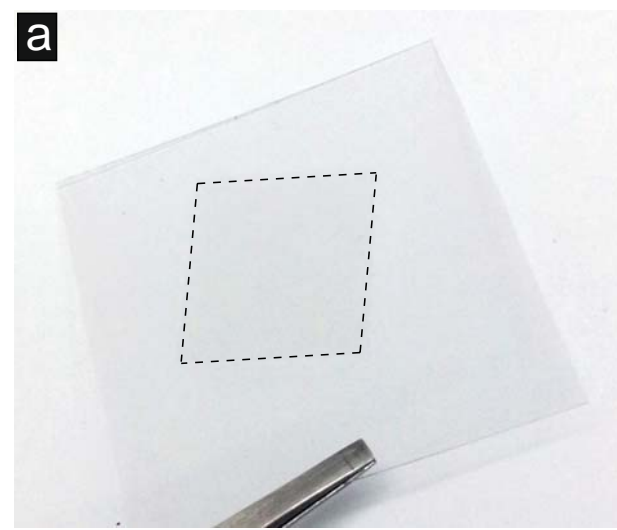
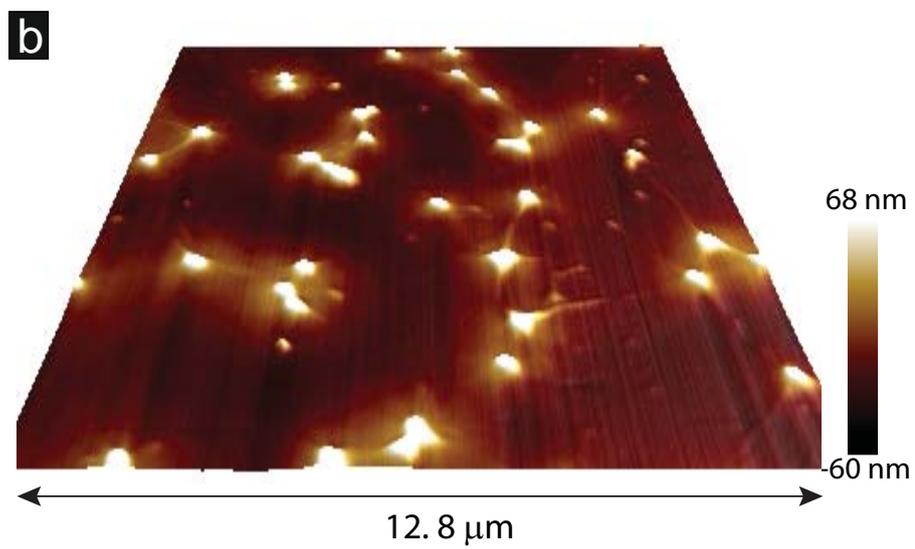
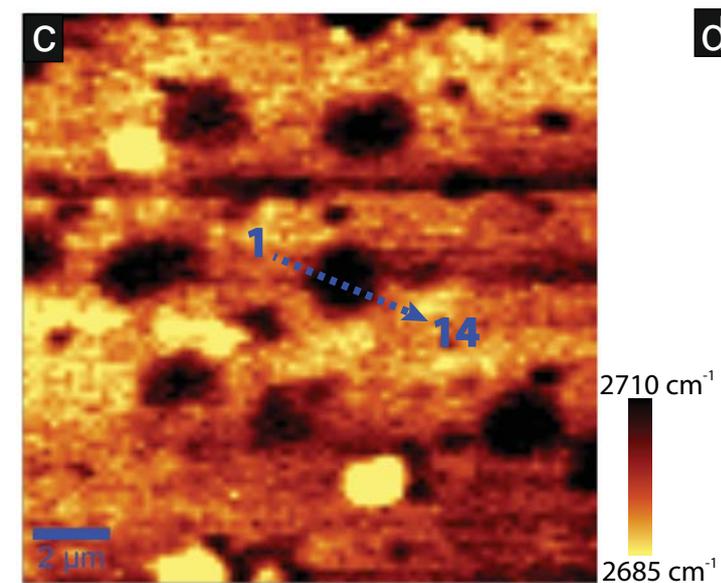
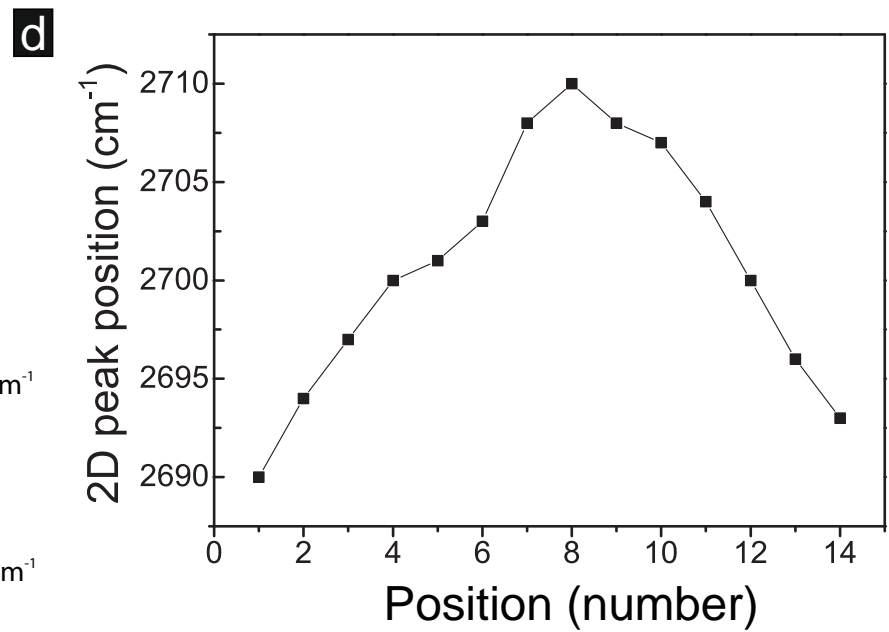

Figure. 1 Ni Guangxin et al.

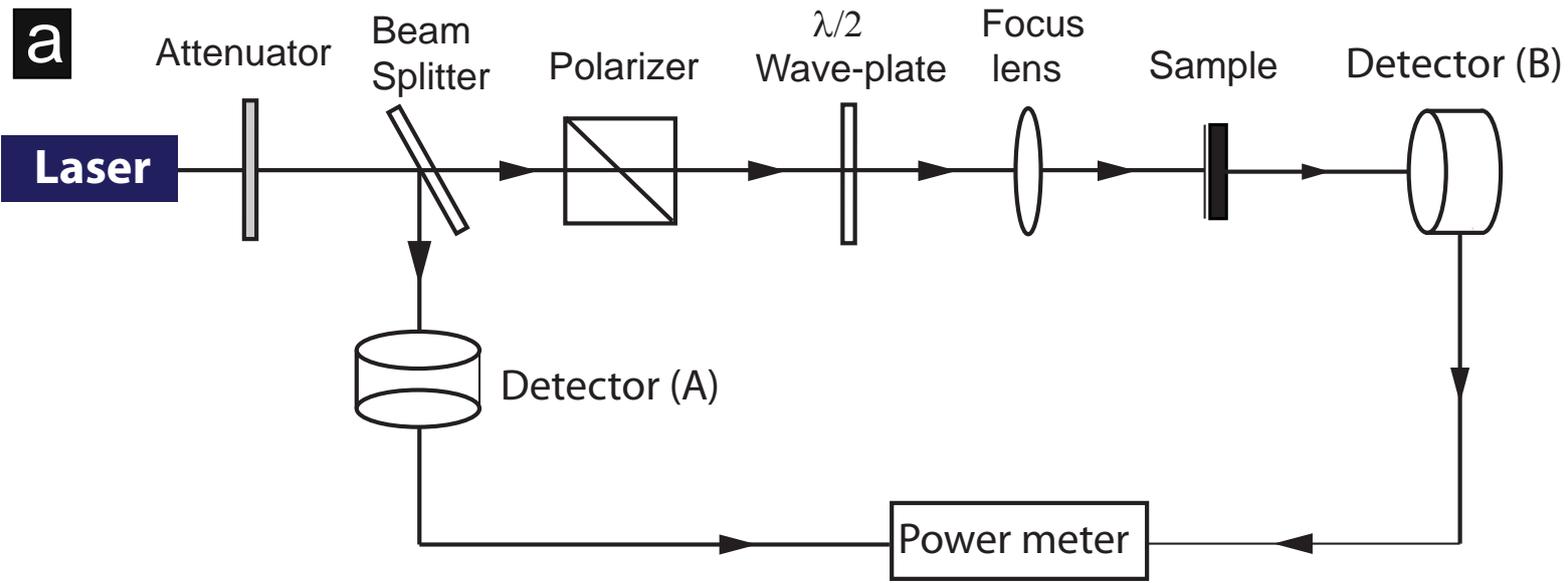

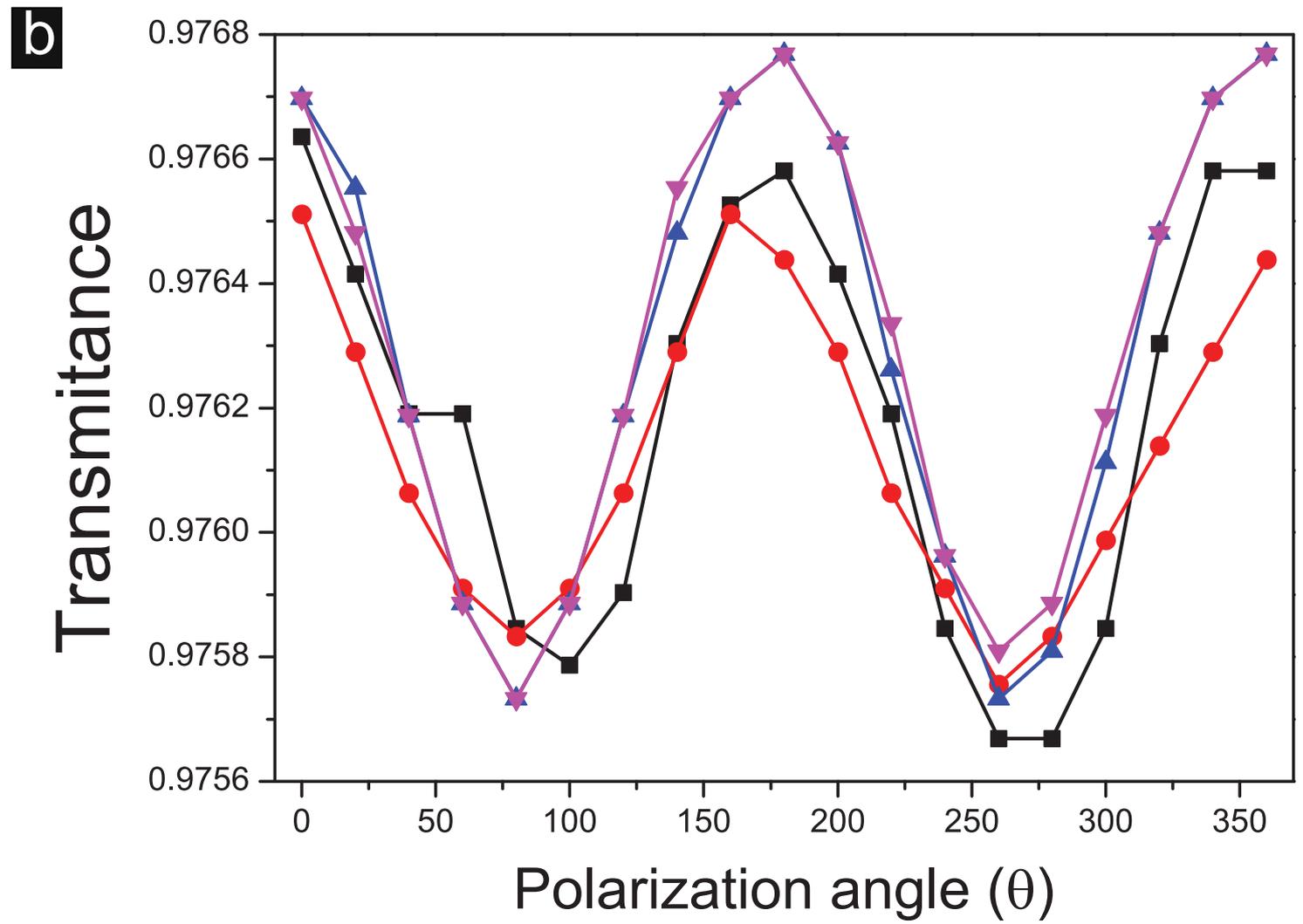

Figure. 2 Ni Guangxin et al.

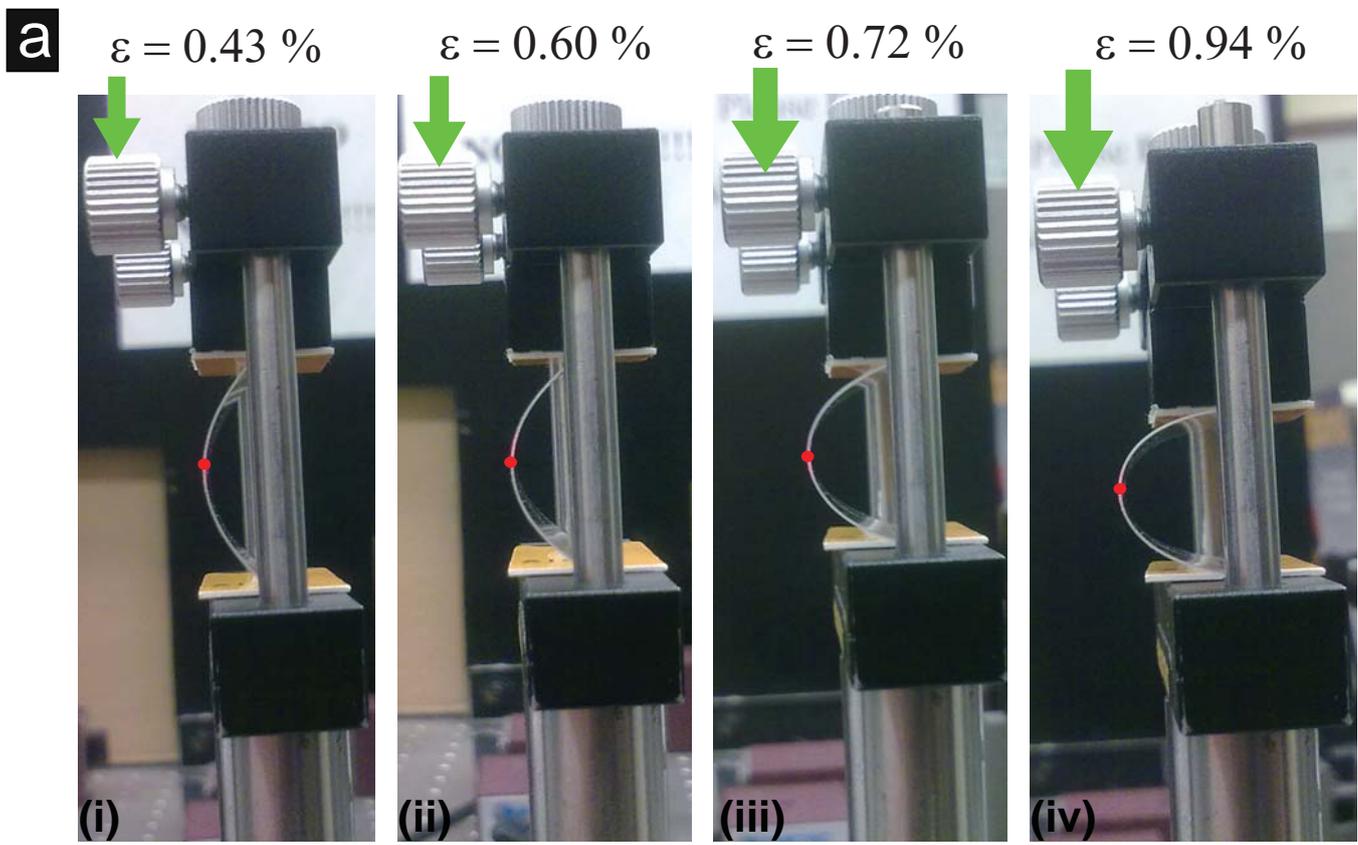

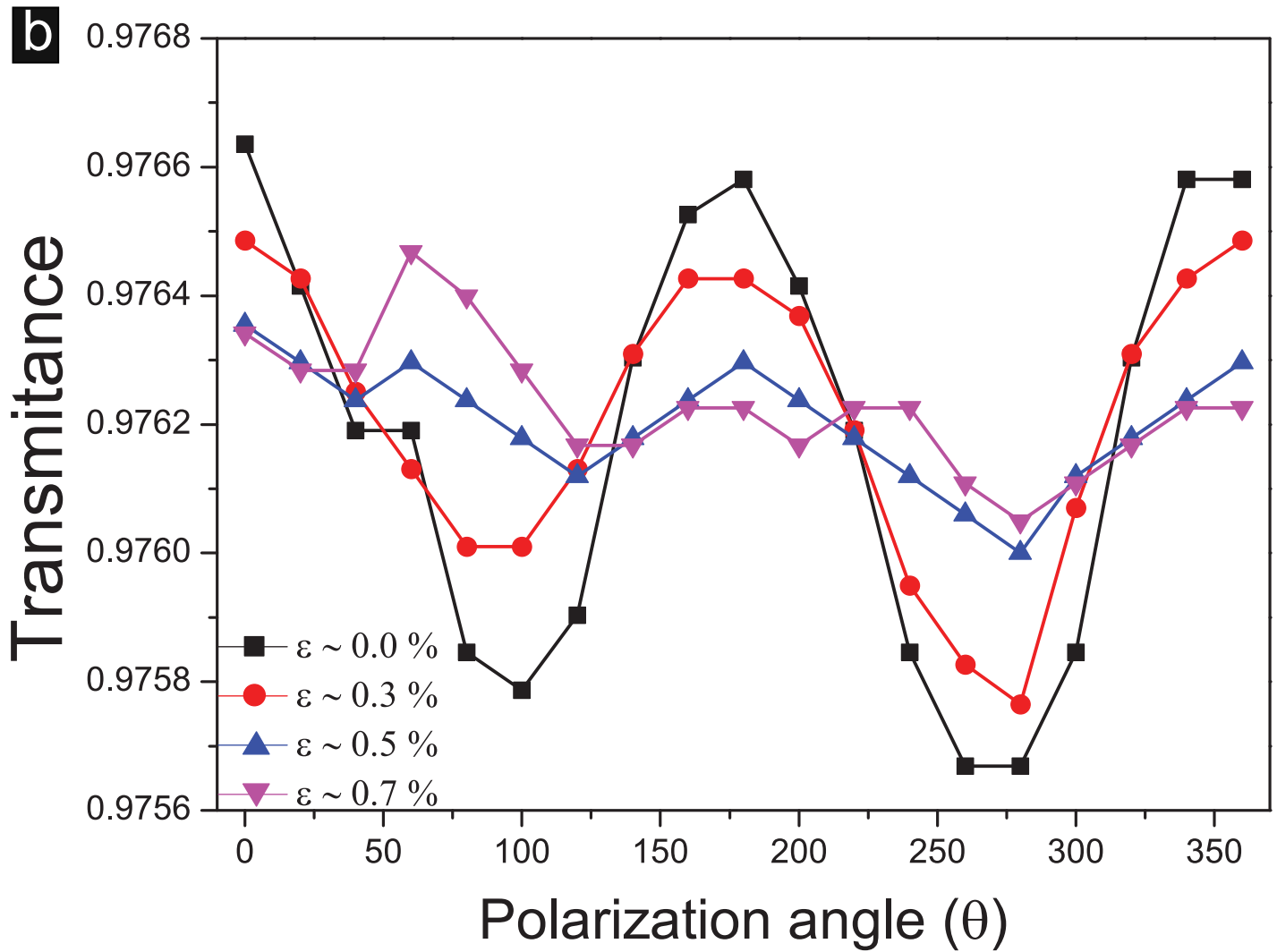

Figure. 3 Ni Guangxin et al.

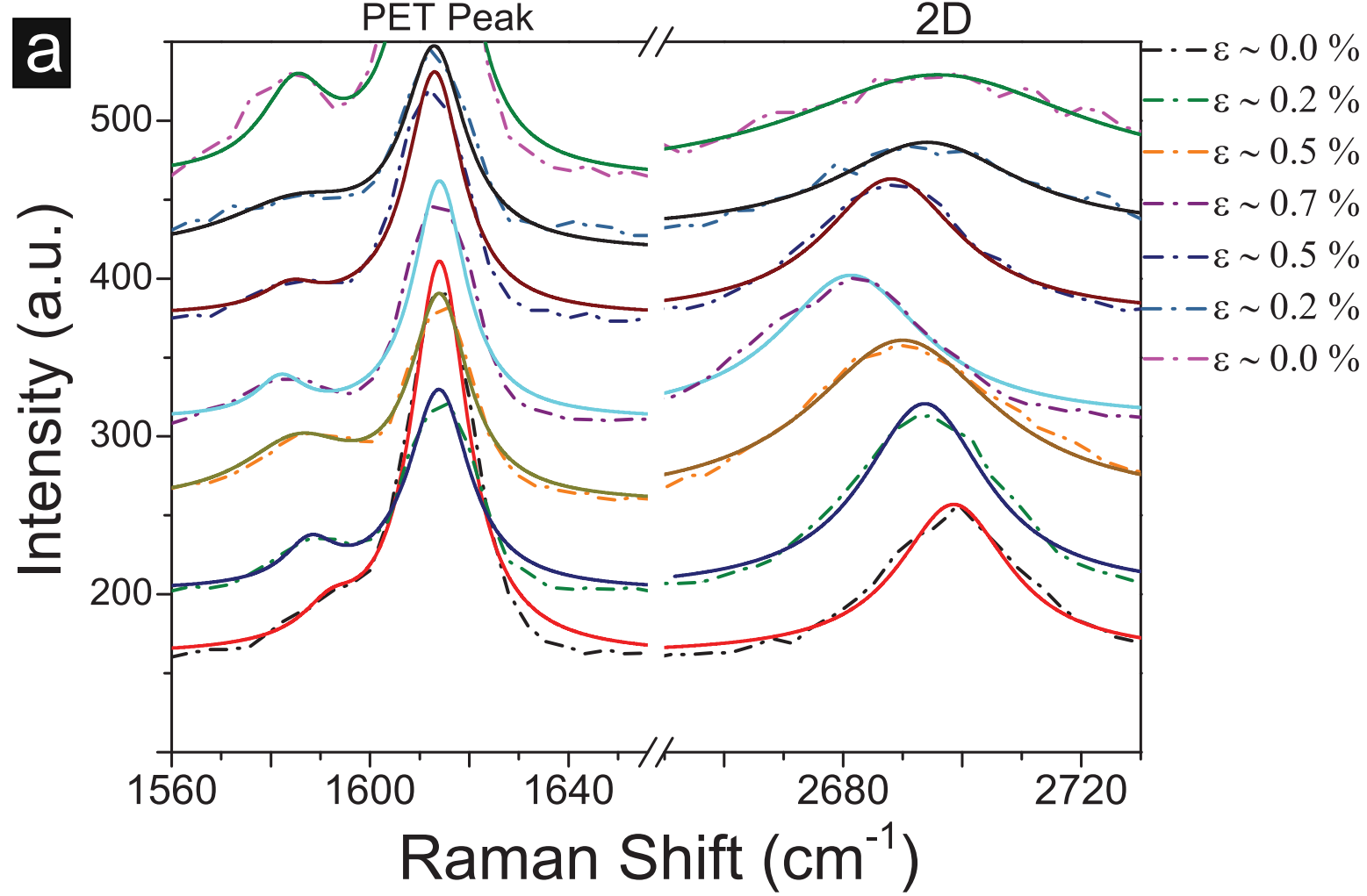
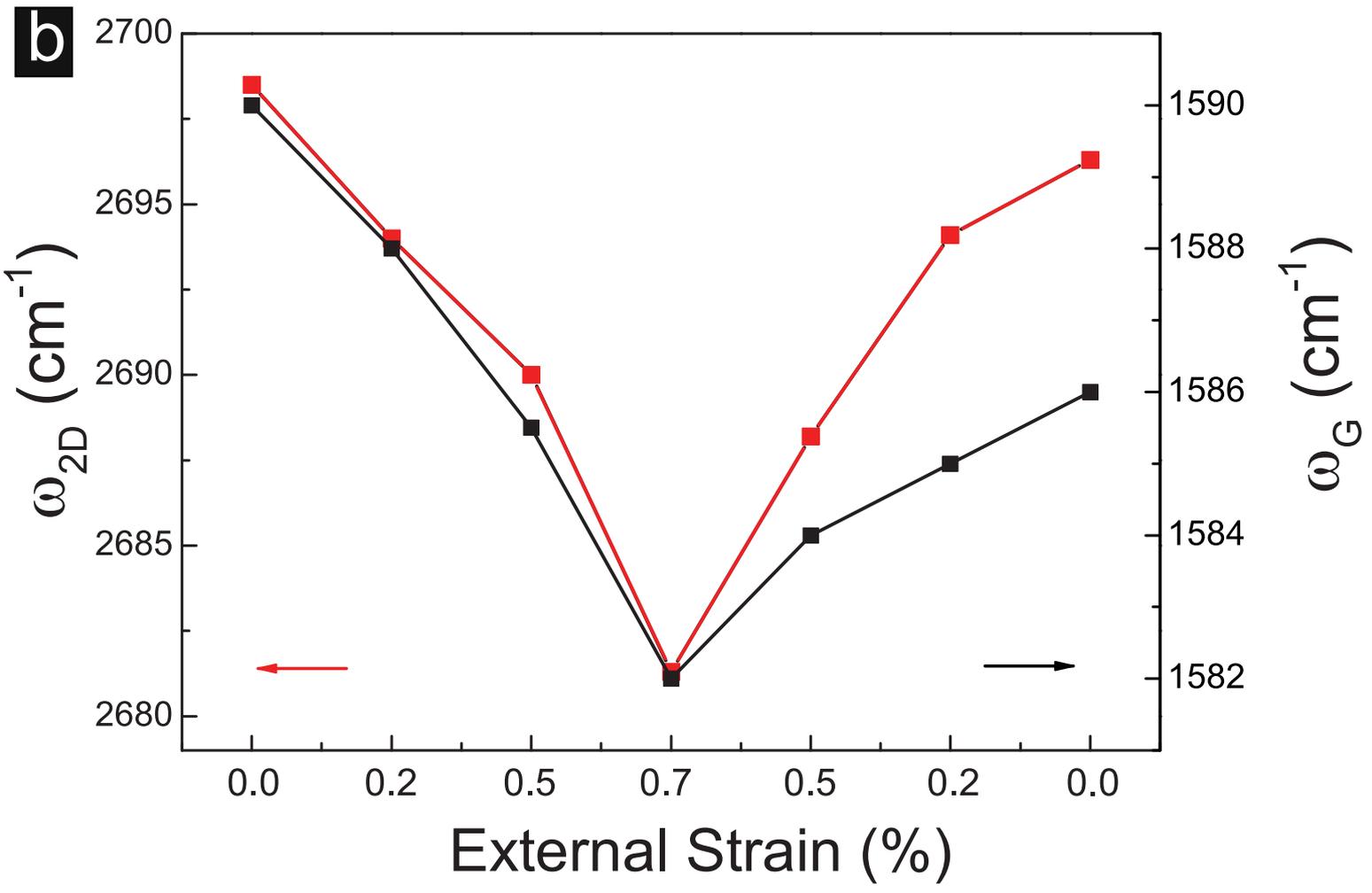

Figure. 4 Ni Guangxin et al.